\documentclass[prl,twocolumn,preprintnumbers,amsmath,amssymb]{revtex4}
\usepackage{epsfig,psfrag}

\catcode`\@=11
\newcommand{\insertfig}[2]{\mbox{\epsfxsize=#1cm \epsfbox{#2.eps}}}
\def \be  {\begin{equation}}
\def \ee  {\end{equation}}
\def \ba  {\begin{eqnarray}}
\def \ea  {\end{eqnarray}}
\def \baa {\begin{eqnarray*}}
\def \eaa {\end{eqnarray*}}
\def \bb  {\begin {thebibliography} }
\def \eb  {\end{thebibliography}}
\def \lab #1 {\label{#1}}
\newcommand{\ft}[2]{{\textstyle\frac{#1}{#2}}}

\newcommand\re[1]{(\ref{#1})}

\def \matrix #1 {\left(\begin{array}{cc} #1 \end{array}\right)}

\newcommand{\as}{\ifmmode\alpha_{\rm s}\else{$\alpha_{\rm s}$}\fi}
\newcommand{\asbar}{\ifmmode\bar{\alpha}_{\rm s}\else{$\bar{\alpha}_{\rm s}$}\fi}

\font\cmss=cmss12 
\def\inbar{\,\vrule height1.5ex width.4pt depth0pt}
\def\IC{\relax\hbox{$\inbar\kern-.3em{\rm C}$}}
\def\IZ{\relax{\hbox{\cmss Z\kern-.4em Z}}}
\def\IR{{\hbox{{\rm I}\kern-.2em\hbox{\rm R}}}}

\def\IP{{\hbox{{\rm I}\kern-.2em\hbox{\rm P}}}}
\def\II{\hbox{{1}\kern-.25em\hbox{l}}}

\begin{document}

\preprint{LPT-Orsay-04-130}

\title{Integrability in Yang-Mills theory on the light cone beyond leading order}

\author{A.V. Belitsky$^1$, G.P. Korchemsky$^2$, D. M\"uller$^3$}

\affiliation{$^1$Department of Physics and Astronomy, Arizona State University, Tempe,
AZ 85287-1504, USA}
\affiliation{$^2$Laboratoire de Physique Th\'eorique, Universit\'e de Paris XI, 91405
Orsay C\'edex, France}
\affiliation{$^3$Institut f\"ur Theoretische Physik, Universit\"at Regensburg, D-93040
Regensburg, Germany}

\begin{abstract}
The one-loop dilatation operator in Yang-Mills theory possesses a hidden
integrability symmetry in the sector of maximal helicity Wilson operators. We
calculate two-loop corrections to the dilatation operator and demonstrate that while
integrability is broken for matter in the fundamental representation of the $SU(3)$
gauge group, for the adjoint $SU(N_c)$ matter it survives the conformal symmetry breaking
and persists in supersymmetric $\mathcal{N}=1$, $\mathcal{N}=2$ and $\mathcal{N}=4$
Yang-Mills theories.
\end{abstract}

\maketitle

Four-dimensional Yang-Mills (YM) theory exhibits a new symmetry which manifests
itself on the quantum level through integrability property of the dilatation
operator in the sector of the maximal helicity Wilson operators in the
multi-color limit \cite{BraDerMan98}. In supersymmetric Yang-Mills (SYM)
theories, integrability is promoted to a larger class of operators and ultimately
to all Wilson operators in the maximally supersymmetric $\mathcal{N} =4$ theory
\cite{BeiSta03}. It is expected that the $\mathcal{N}=4$ dilatation operator has
to be integrable to all orders in 't Hooft coupling since the $\mathcal{N} = 4$
SYM is dual to the superstring theory on AdS$_5 \times$S$^5$ background
\cite{Mal97}, whose world-sheet sigma-model possesses an infinite number of
integrals of motion both on classical \cite{BenPolRoi03} and quantum levels
\cite{Ber04}. So far integrability in YM theory was unequivocally established at
leading order in coupling and it was argued to hold in certain closed subsectors
of the $\mathcal{N} = 4$ SYM theory in higher loops \cite{BeiKriSta03}. In
distinction with the $\mathcal{N}=4$ model, the conformal symmetry of YM theory
is broken by quantum corrections and the conformal anomaly affects the dilatation
operator starting from two loops. A natural question arises whether integrability
in the sector of the maximal helicity operators carries on to higher orders in YM
theory and its supersymmetric extensions.

In this Letter, we report on a calculation of the two-loop dilatation operator in the sector
of three-quark (baryon) operators of the maximal helicity-$3/2$. Integrability is a genuine
symmetry of YM theories and the choice of the sector is driven solely by the simplicity of
the computations involved. For our purposes we consider separately the case of quarks in the
fundamental $SU(3)$ and adjoint $SU(N_c)$ representations. The former corresponds to QCD while
the latter is relevant to its supersymmetric extension. The corresponding three-quark
light-cone operators are defined as
\ba
\mathbb{B}_{\rm f} (z_1, z_2, z_3) \!\!\!&=&\!\!\!
\varepsilon^{ijk} q^\uparrow_{i} (z_1) q^\uparrow_{j} (z_2) q^\uparrow_{k} (z_3)
\, , \nonumber\\
\mathbb{B}_{\rm a}(z_1, z_2, z_3) \!\!\!&=&\!\!\! {\rm tr} \, [ q^\uparrow (z_1) q^\uparrow (z_2)
q^\uparrow (z_3) ] \, ,
\label{B-operator}
\ea
where $q^\uparrow(z) = \frac12 (1 - \gamma_5) q(z n_\mu)$ is the helicity-$1/2$
fermion ``living'' on the light-cone ($n_\mu^2 = 0$) and, in the second relation,
$q = q^a t^a$ with the $SU (N_c)$ generators $t^a$ in the fundamental
representation. The quarks in \re{B-operator} may have different flavors and
their total number is $N_f$. It is tacitly assumed that the gauge invariance is
restored in \re{B-operator} by inserting appropriate Wilson lines between quark
fields. Later on we shall adopt the light-like axial gauge $n \cdot A = 0$ in
which the gauge links reduce to the unit matrix. The light-cone operators
\re{B-operator} are generating functions of local Wilson operators. The latter
can be obtained by Taylor expanding $\mathbb{B}(z_i)$ in the light-cone
separations $z_i$. The operator $\mathbb{B}_{\rm f} (z_i)$ has a direct
phenomenological interest in QCD as its matrix element defines the
$\Delta^{3/2}-$distribution amplitude \cite{BroLep80}. For $N_f = 1$, the
operator $\mathbb{B}_{\rm a} (z_i)$ is a component of a supermultiplet in the
$\mathcal{N} = 1$ SYM theory. Its remaining components are obtained from
$\mathbb{B}_{\rm a}(z_i)$ by replacing $q^\uparrow$ one by one with helicity-up
gluon fields, thus covering the maximal-helicity three-gluon operator.

\begin{figure*}[t]
\begin{center}
\mbox{
\begin{picture}(0,42)(200,0)
\put(0,-8){\insertfig{14}{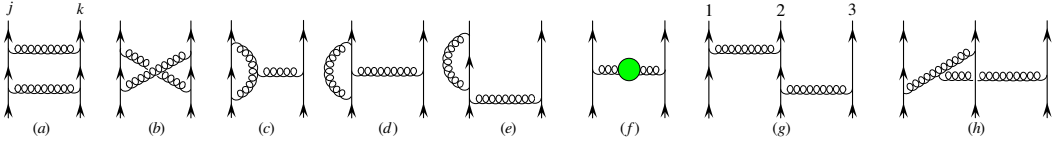}}
\end{picture}
}
\end{center}
\caption{\label{twoloopkernel} Two-loop corrections to the two-particle ($a-f$)
and three-particle ($g,h$) evolution kernels in \re{H-2-loop}.}
\end{figure*}

The goal of this study is to elucidate the symmetry properties of the dilatation
operator governing the scale dependence of the operators (\ref{B-operator}) to
two-loop order. A unique feature of the operators (\ref{B-operator}) is that they
mix only with themselves and obey the Callan-Symanzik equation
\be
\label{DilatationOperator}
\left( \mu \frac{\partial}{\partial \mu} + \beta (g)
\frac{\partial}{\partial g}
\right) \mathbb{B}(z_i) = -
[\mathbb{H} \cdot \mathbb{B}](z_i) \, .
\ee
Here $\mathbb{H}$ is an integral operator acting on light-cone coordinates of quark fields
and defining a representation of the dilatation operator $\mathbb{D}$ in the space spanned
by light-cone operators, $i [ \mathbb{D} ,\mathbb{B}(z_i)] = (\mathbb{H} + 3 d_q)\mathbb{B}
(z_i)$ with $d_q=3/2$ the quark canonical dimension. Its perturbative expansion
\be
\label{TwoLoopHamiltonian}
\mathbb{H}
=
\lambda \, \mathbb{H}^{\scriptscriptstyle (0)}_{\phantom{i}}
+
\lambda^2 \, \mathbb{H}^{\scriptscriptstyle (1)}_{\phantom{i}}
+
\mathcal{O} (\lambda^3)
\, ,
\ee
goes in $\lambda$, depending on the quark representation in \re{B-operator},
\be
\label{CouplingConstant}
\lambda_{\rm f} = \frac{g^2}{8 \pi^2} \frac{4}{3}
\, , \qquad
\lambda_{\rm a} = \frac{g^2}{8 \pi^2} N_c
\, ,
\ee
and gauge coupling $g$. Notice that the leading order evolution kernel
$\mathbb{H}^{\scriptscriptstyle (0)}_{\phantom{i}}$ is universal (see \re{LOkernel}
below) while the two-loop kernel $\mathbb{H}^{\scriptscriptstyle (1)}_{\phantom{i}}$
is different for $\mathbb{B}_{\rm f}(z_i)$ and $\mathbb{B}_{\rm a}(z_i)$.

To solve \re{DilatationOperator} one has to examine the Schr\"odinger equation for the evolution
kernel \re{TwoLoopHamiltonian}
\be
\label{Schrodinger}
[ \mathbb{H} \cdot \Psi_{q} ] (z_1, z_2, z_3) = \gamma_{q} (\lambda) \,
\Psi_{q} (z_1, z_2, z_3) \, ,
\ee
with eigenvalues $\gamma_q (\lambda)$ having a perturbative expansion in the coupling constants
\re{CouplingConstant} similar to Eq.\ \re{TwoLoopHamiltonian}, and
\be
\Psi_q(z_i) = \Psi_q^{\scriptscriptstyle (0)}(z_i) + \lambda \,
\Psi_q^{\scriptscriptstyle (1)}(z_i) + \mathcal{O} (\lambda^2)
\label{Psi}
\ee
being homogeneous polynomials of degree $N=0,1,\ldots$. Having solved \re{Schrodinger}, one can
expand nonlocal operators $\mathbb{B}(z_i)$ over local Wilson operators $\mathbb{O}_q(0)$
built from three quark fields and $N$ covariant derivatives acting on them
\be
\mathbb{B} (z_1, z_2, z_3) = \sum\nolimits_{q} \Psi_q (z_1, z_2, z_3) \,
\mathbb{O}_q (0)
\, .
\ee
To one-loop order, these operators have autonomous scale dependence but this property is lost in
two loops due to dependence of $\Psi_q(z_i)$ on the coupling constant, Eq.~\re{Psi}. Integrability
arises as a hidden symmetry of the Schr\"odinger equation \re{Schrodinger} through the existence
of a new quantum number $q$.

To one-loop order, the kernel $\mathbb{H}^{\scriptscriptstyle (0)}_{\phantom{i}}$ has a pair-wise
form
\be
\label{LOkernel} \mathbb{H}^{\scriptscriptstyle (0)}_{\phantom{i}}
=
\mathbb{H}_{12}^{\scriptscriptstyle (0)}
+
\mathbb{H}_{23}^{\scriptscriptstyle (0)}
+
\mathbb{H}_{31}^{\scriptscriptstyle (0)} + \ft32 \, ,
\ee
with the two-particle kernel $\mathbb{H}_{jk}^{(0)}$ displacing $j^{\rm th}$
and $k^{\rm th}$ quarks along the light-cone in the direction of each other
\ba
\label{LightConeKernel}
&&\!\!\!\!\!\!
[ \mathbb{H}_{12}^{\scriptscriptstyle (0)} \cdot \mathbb{B} ] (z_1, z_2, z_3)
=
\int_0^1 \frac{d \alpha}{\alpha} \bar\alpha^{2 j_q - 1}
{}\Big[ 2 \mathbb{B} (z_1, z_2, z_3)
\\
&&\qquad\quad \ \ - \mathbb{B} (\bar\alpha z_1 + \alpha z_2, z_2, z_3) -
\mathbb{B} (z_1, \alpha z_1 + \bar\alpha z_2, z_3) {}\Big], \nonumber
\ea
where $\bar\alpha \equiv 1 - \alpha$ and $j_q = 1$ is the conformal spin of the
quark. The one-loop dilatation operator \re{LOkernel} possesses two conserved
charges $q_2^{\scriptscriptstyle (0)}, q_3^{\scriptscriptstyle (0)}$ such that
$[\mathbb{H}^{\scriptscriptstyle (0)}_{\phantom{i}}, q_i^{\scriptscriptstyle (0)}]
= 0$,
\be
q_2^{\scriptscriptstyle (0)} = L_{12}^2 + L_{23}^2 + L_{31}^2 \, , \qquad
q_3^{\scriptscriptstyle (0)} = i[L_{12}^2, L_{23}^2] \, ,
\label{charges}
\ee
where $L_{jk}^2$ is a two-particle Casimir operator of the ``collinear'' $SL(2)$
subgroup of the conformal group
\be
L_{jk}^2\, \mathbb{B} (z_i) = -
z_{jk}^{2(1-j_q)}\partial_{z_j}\partial_{z_k}z_{jk}^{2j_q}\,\mathbb{B} (z_i)\,,
\label{L2}
\ee
with $z_{jk}=z_j-z_k$. Complete integrability implies that the one-loop
dilatation operator is a function of the conserved charges
$\mathbb{H}^{\scriptscriptstyle (0)}_{\phantom{i}} =
\mathbb{H}^{\scriptscriptstyle (0)}_{\phantom{i}} (q_2^{\scriptscriptstyle
(0)}, q_3^{\scriptscriptstyle (0)})$. So that its eigenfunctions diagonalize
both of them simultaneously, e.g., $q_3^{\scriptscriptstyle (0)}
\Psi_q^{\scriptscriptstyle (0)} (z_i) = q \Psi_q^{\scriptscriptstyle (0)}
(z_i)$. Remarkably, $\mathbb{H}^{\scriptscriptstyle (0)}_{\phantom{i}}$ can be
mapped into the $SL(2,\mathbb{R})$ Heisenberg magnet of spin$-j_q$ and length
equal to the number of quarks \cite{BraDerMan98}. As a result, the
Schr\"odinger equation (\ref{Schrodinger}) is completely integrable to one-loop
order and is solved exactly using the Bethe Ansatz.

The one-loop integrability holds irrespective of the $SU(N_c)$ representation
of the quark fields in \re{B-operator}. To verify whether this symmetry is
preserved beyond leading order, we calculate the two-loop dilatation operators
\re{TwoLoopHamiltonian} in the light-like axial gauge $n \cdot A = 0$. In this
gauge, the relevant Feynman diagrams are shown in Fig.\ \ref{twoloopkernel}.
Their contribution to the two evolution kernels $\mathbb{H}^{\scriptscriptstyle
(1)}_{\rm f}$ and $\mathbb{H}^{\scriptscriptstyle (1)}_{\rm a}$ only differs by
the accompanying color factors. As a consequence, the two kernels are different
but have the same general form
\be
\mathbb{H}^{\scriptscriptstyle (1)}_{\phantom{i}} = \mathbb{H}_{12} +
\mathbb{H}_{23} + \mathbb{H}_{31} + \mathbb{H}_{123}+
\Gamma^{\scriptscriptstyle (1)} \, .
\label{H-2-loop}
\ee
Here $\mathbb{H}_{jk}$ is the two-particle kernel computed from the graphs in
Fig.\ \ref{twoloopkernel} ($a$) -- ($f$) and $\mathbb{H}_{123}$ is the
three-particle irreducible kernel defined by the diagram ($g$) and its cyclic
permutations. The diagram ($h$) vanishes for both operators in \re{B-operator}.
Notice that the nonplanar diagram ($b$) contributes to $\mathbb{B}_{\rm f}(z_i)$ and
vanishes for $\mathbb{B}_{\rm a}(z_i)$. The quark-leg renormalization (not displayed)
contribute to the c-number constant $\Gamma^{\scriptscriptstyle (1)}$.

By analogy with one-loop, Eq.~\re{LOkernel}, we define the two-loop Hamiltonians
$\mathbb{H}_{jk}$ and $\mathbb{H}_{123}$ in such a way that they annihilate the
local baryon operator $\mathbb{B}_{\rm a,f} (z_i=0)$ so that its anomalous
dimension equals $[\mathbb{H}-\gamma_0 (\lambda)]\mathbb{B}(z_i=0)=0$ with
$\gamma_0 (\lambda) = \lambda \, \frac{3}{2} + \lambda^2 \,
\Gamma^{\scriptscriptstyle (1)} + \mathcal{O} (\lambda^3)$. To save space we do
not present the explicit expression for $\Gamma^{\scriptscriptstyle (1)}$. It
defines the ground state energy for the Hamiltonian $\mathbb{H}$,
Eq.~\re{Schrodinger}, and is not relevant for the discussion of integrability.

The diagram in Fig.\ \ref{twoloopkernel} ($f$) involves a quark loop. Its
contribution is proportional to the number of flavors $N_f$ and is gauge
invariant. This allows one to split the two-particle kernel $\mathbb{H}_{jk}$
into the sum of gauge-invariant terms
\be
\label{BetaInKernel}
\mathbb{H}_{jk} = \mathbb{H}_{jk}^{(\beta_0=0)} +
 ({\beta_{0}}/{c_{\scriptscriptstyle\rm R }}) \,
\mathbb{H}_{jk}^{\scriptscriptstyle (1)}
\, ,
\ee
where the second one is proportional to the one-loop $\beta$-function
$\beta_0 = 11 N_c/3 -4T_{\scriptscriptstyle\rm R } N_f/3$ in the underlying
YM theory with $T_{\rm f} = 1/2$, $c_{\rm f}=2/3$ and $T_{\rm a} = c_{\rm a} =N_c/2$.

Starting from two loops the anomalous dimensions receive conformal symmetry
breaking corrections due to nonzero $\beta$-function \re{BetaInKernel} and, in
addition, they depend on the scheme used to subtract ultraviolet divergences. In
our calculations we applied the dimensional regularization with
$d=4-2\varepsilon$ and subtracted poles in $1/\varepsilon$ in the $\overline{\rm
MS}-$scheme. In general, supersymmetry is broken in the dimensional
regularization and is preserved in the dimensional reduction. The diagrams in
Fig.\ \ref{twoloopkernel} give identical results for both procedures so that
supersymmetry is preserved. The difference arises only for two-loop
quark self-energy, which contributes to $\Gamma^{\scriptscriptstyle (1)}$. The
resulting expressions for the two-loop Hamiltonians $\mathbb{H}_{jk}$ and
$\mathbb{H}_{123}$, Eq.~\re{H-2-loop}, are cumbersome and will be given
elsewhere.

The dilatation operator $\mathbb{H}$, Eq.~\re{TwoLoopHamiltonian}, is invariant
under the cyclic permutation of quarks $\mathbb{P}$ and the interchange of any
pair of quarks $\mathbb{P}_{jk}$. Since $[\mathbb{P},\mathbb{P}_{jk}]\neq 0$,
the eigenfunctions $\Psi_q(z_i)$ cannot diagonalize both operators
simultaneously. Choosing $\mathbb{P}$, the eigenstates can be classified with
respect to the quasimomentum
\be
\mathbb{P} \, \Psi_q (z_1, z_2, z_3) =
\Psi_q (z_2, z_3, z_1) = {\rm e}^{i \theta_q} \Psi_q (z_1, z_2, z_3)
\ee
with $\theta_q = 2 \pi n/3$ and $n=-1,0,1$. The parity symmetry of the
Hamiltonian, $[\mathbb{P}_{12}, \mathbb{H}]=0$, combined with the identity
$\mathbb{P} \, \mathbb{P}_{12} = \mathbb{P}_{12} \, \mathbb{P}^2$ implies that,
independently of the existence of the charge $q_3$, the eigenvalues of
$\mathbb{H}$ with non-zero quasimomentum $\theta_q\neq 0$ are double degenerated,
\be
[\mathbb{H}-\gamma_q(\lambda)]\Psi_q^\pm(z_i)=0\,,\quad \Psi_q^\pm=\ft12{(1\pm
\mathbb{P}_{12})}\Psi_q(z_i),
\label{parity}
\ee
whereas for $\theta_q=0$ the eigenvalues are not necessarily degenerated. The
existence of the conserved charge $q_3=q_3^{\scriptscriptstyle
(0)}+\lambda\,q_3^{\scriptscriptstyle (1)}+ ...$ symmetric with respect to
$\mathbb{P}$ but antisymmetric with respect to $\mathbb{P}_{12}$, i.e.,
$[\mathbb{P}, q_3] = \{ \mathbb{P}_{12}, q_3\} = 0$, extends the degeneracy
property to the eigenstates with $\theta_q = 0$ and $q_3\neq 0$, since
$\mathbb{P}_{12}\, \mathbb{H}(q_3)\, \mathbb{P}_{12} = \mathbb{H}(- q_3)$.

At one-loop, integrability of the Schr\"odinger equation \re{Schrodinger} implies double
degeneracy of all eig\-en\-values of $\mathbb{H}^{\scriptscriptstyle (0)}$, except those
with $q_3=0$. To determine two-loop corrections to the eigenspectrum of $\mathbb{H}$,
Eqs.~\re{H-2-loop}, we apply the conventional `degenerate perturbation theory'. Using
the basis of one-loop definite-parity eigenstates $\Psi_q^{\scriptscriptstyle(0)\pm}
(z_i)$, Eq.~\re{parity}, normalized with respect to the $SL(2;\mathbb{R})$
invariant scalar product as $\langle \Psi_{q}^{\scriptscriptstyle (0)\pm} |
\Psi_{q'}^{\scriptscriptstyle (0)\pm} \rangle=\delta_{qq'}$, one finds
\ba
\gamma^{\scriptscriptstyle (1)\pm}_q &=& \langle \Psi_{q}^{\scriptscriptstyle
(0) \pm} | \mathbb{H}^{\scriptscriptstyle (1)}_{\phantom{i}} |
\Psi_{q}^{\scriptscriptstyle (0) \pm} \rangle \, ,
\label{PT}
\\
\Psi_q^{\scriptscriptstyle (1)\pm} &=& \sum_{q' \neq q} \frac{ \langle
\Psi_{q'}^{\scriptscriptstyle (0)\pm}| \mathbb{H}^{\scriptscriptstyle
(1)}_{\phantom{i}} | \Psi_{q}^{\scriptscriptstyle (0)\pm} \rangle }{
\gamma^{\scriptscriptstyle (0)}_q - \gamma^{\scriptscriptstyle (0)}_{q'} }
\Psi_{q'}^{\scriptscriptstyle (0)\pm} \equiv \mathbb{Z}^\pm\,
\Psi_q^{\scriptscriptstyle (0)\pm} \, . \nonumber
\ea
Going over through lengthy calculations we observed that, in agreement with our
expectations, the eigenstates with nonzero quasimomentum $\theta_q\neq 0$ are
double degenerated to two loops for both operators in \re{B-operator}. For the
eigenstates with $\theta_q=0$ the situation is quite different. We found that the
desired pairing of eigenvalues occurs only for the operator $\mathbb{B}_{\rm
a}(z_i)$ while it is lifted for the $\mathbb{B}_{\rm f}(z_i)$.

\begin{figure*}[t]
\begin{center}
\mbox{
\begin{picture}(0,105)(200,0)
\psfrag{N}[cc][cc]{$\scriptscriptstyle N$}
\psfrag{(a)}[cc][cc]{$(a)$}\psfrag{(b)}[cc][cc]{$(b)$}\psfrag{(c)}[cc][cc]{$(c)$}
\psfrag{Delta0}[cc][cc]{$\gamma_N^{\scriptscriptstyle (0)}$}
\psfrag{Delta1}[cc][cc]{$\gamma_N^{\scriptscriptstyle (1)}$}
\put(-20,-10){\insertfig{4.5}{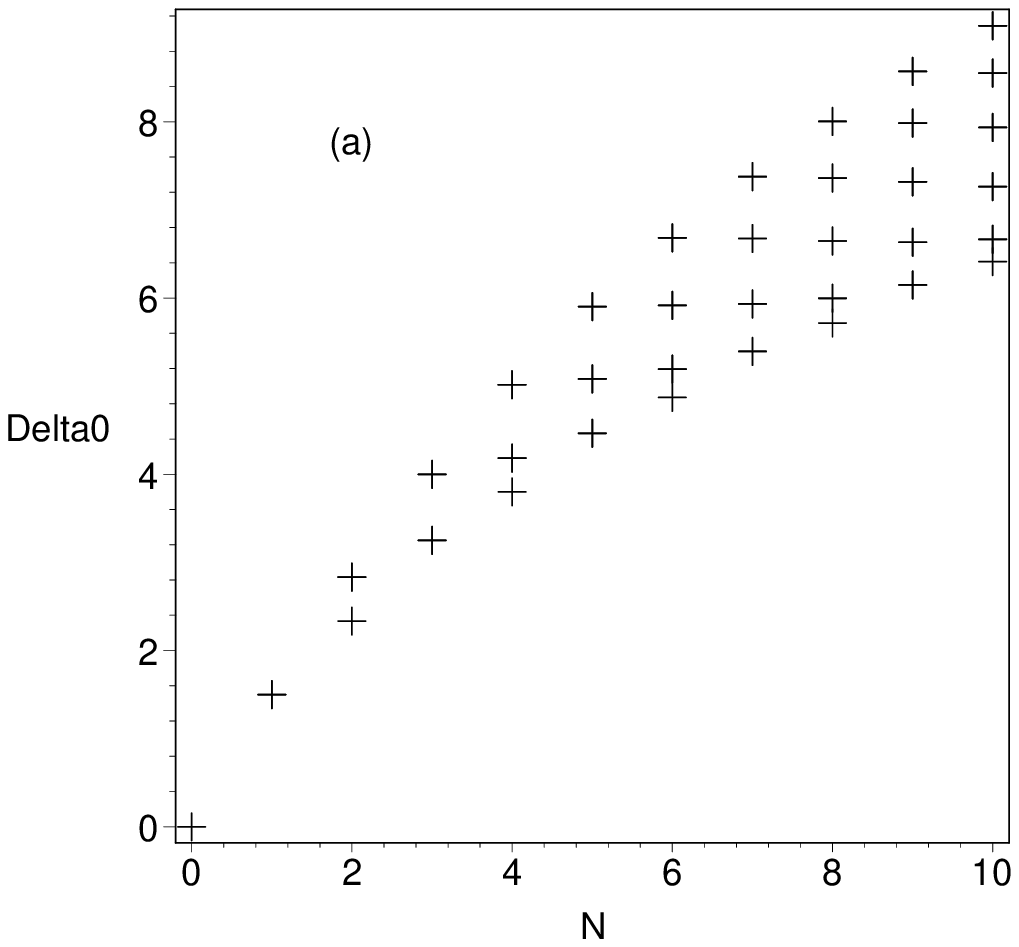}}
\put(140,-10){\insertfig{4.5}{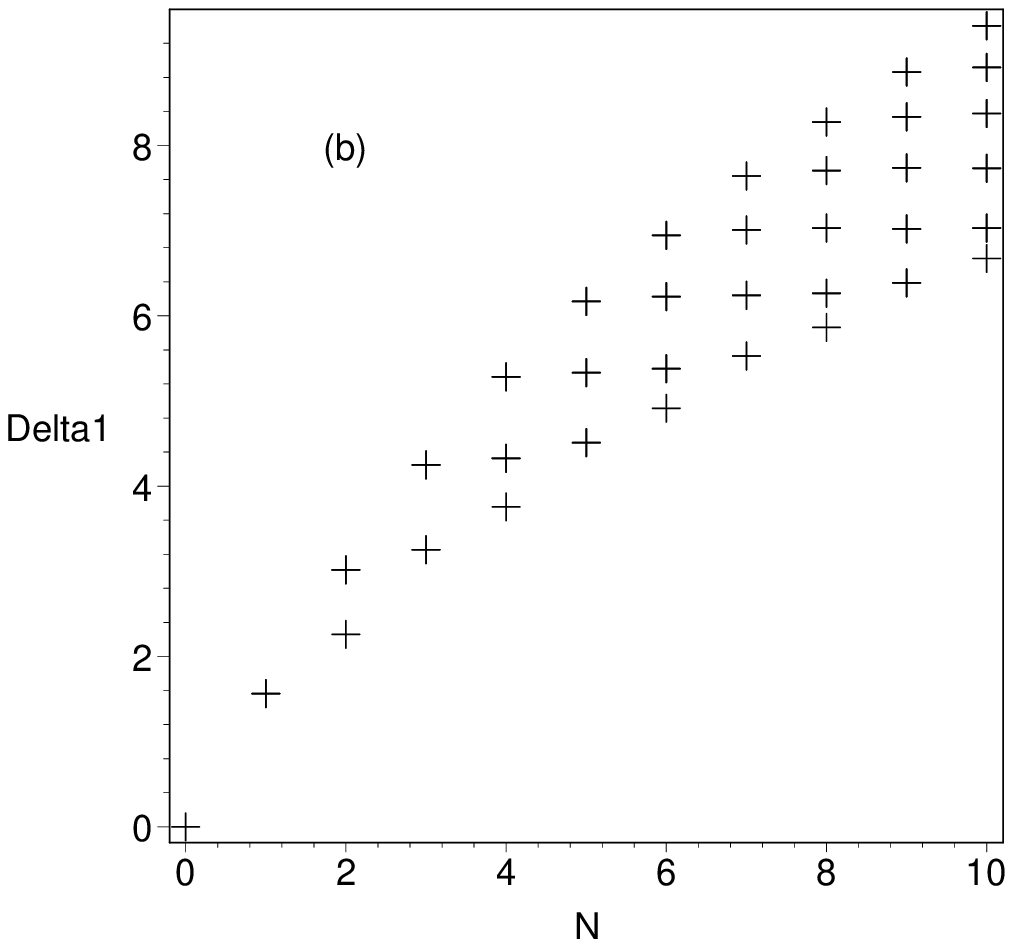}}
\put(300,-10){\insertfig{4.5}{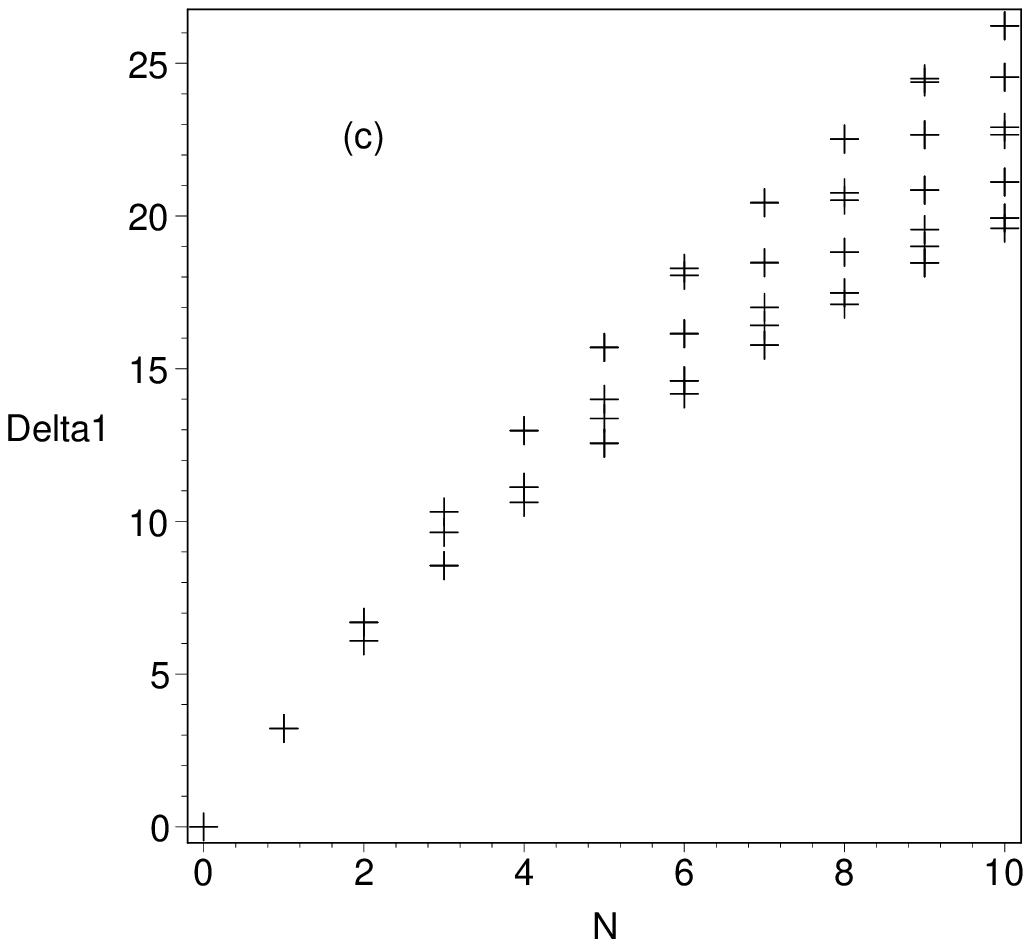}}
\end{picture}
}
\end{center}
\caption{\label{SpectrumAD} Spectrum of anomalous dimensions (see text) at one
loop ($a$), and two loops for quarks in the adjoint ($b$) and fundamental ($c$)
representations. In the latter case, more states for the same conformal spin $N$
arise due to lifting of the degeneracy.}
\end{figure*}

We remind that the eigenstates $\Psi_q^{\pm}(z_i)$ are homogeneous polynomials
of degree $N=0,1,...$ and the total number of eigenvalues equals $N+1$. The
first non-trial example arises for $N = 3$ and demonstrates the main trend,
recurring for higher $N$. At $N=3$, for quarks in the adjoint representation
one finds two pairs of the eigenvalues $\Delta \gamma^\pm \equiv
\gamma_q^\pm(\lambda)-\gamma_0(\lambda)$,
\ba
\Delta \gamma_{\rm I}^\pm (\lambda) \!\!\!&=&\!\!\!  \lambda_{\rm a}\, 4  +
\lambda_{\rm a}^2 \left[ -3+\ft{29}{24}\ft{\beta_0}{c_{\rm a}}\right],
\label{pair1}
\\
\Delta \gamma_{\rm II}^\pm (\lambda)  \!\!\!&=&\!\!\!  \lambda_{\rm a}\ft{13}{4}+
 \lambda_{\rm a}^2\left[-\ft{1139}{384} + \ft{199}{192}\ft{\beta_0}{c_{\rm a}}
\right], \nonumber
\ea
with $\Delta \gamma_{\rm I}^\pm (\lambda)$ and $\Delta \gamma_{\rm II}^\pm
(\lambda)$ corresponding to the eigenstates with $\theta_q=0$ and $\theta_q=\pm
2\pi/3$, respectively, and ${\beta_0}/{c_{\rm a}}=\ft23(11-2N_f)$. For quarks
in the fundamental representation, the two eigenstates with $\theta_q = 0$ have
different eigenvalues starting from two loops, $\Delta \gamma_{\rm I}^+ \neq
\Delta \gamma_{\rm I}^- $, and we get instead
\ba
\Delta \gamma_{\rm I}^+ (\lambda) \!\!\!&=&\!\!\! \lambda_{\rm f}\, 4 +
\lambda_{\rm f}^2 \left[ -\ft{101}{12} +  \ft{29}{24}\ft{\beta_0}{c_{\rm f}}
\right],
\nonumber \\
\Delta \gamma_{\rm I}^- (\lambda)  \!\!\!&=&\!\!\!  \lambda_{\rm f}\, 4 +
\lambda_{\rm f}^2 \left[ -\ft{291}{32} +  \ft{29}{24}\ft{\beta_0}{c_{\rm f}}
\right],
\label{pair2} \\
\Delta \gamma_{\rm II}^\pm (\lambda) \!\!\!&=&\!\!\!   \lambda_{\rm f}\ft{13}{4}
+ \lambda_{\rm f}^2 \left[ -\ft{721}{96} + \ft{199}{192}\ft{\beta_0}{c_{\rm f}}
\right], \nonumber
\ea
with ${\beta_0}/{c_{\rm f}}=\ft{33}2-N_f$. For $N_f=1$ the exact spectrum of
eigenvalues $\Delta \gamma(\lambda)=\lambda \gamma_N^{\scriptscriptstyle (0)} +
\lambda^2 \gamma_N^{\scriptscriptstyle (1)}$ for the conformal spin $0 \le N \le
10$ is displayed in Fig.\ \ref{SpectrumAD}. It clearly demonstrates the lifting
of the degeneracy and proliferation of eigenvalues for the baryon operator
$\mathbb{B}_{\rm f}(z_i)$ compared to $\mathbb{B}_{\rm a}(z_i)$ where the pairing
persists for all eigenstates.

We recall that the conformal symmetry is broken for $\beta_0\neq 0$. One
immediately sees from \re{pair1} and \re{pair2} that though the degeneracy
is lifted for the operator $\mathbb{B}_{\rm f}$ even in the conformal limit
$\beta_0=0$, the $\beta_0$-terms themselves do preserve the ``pairing" of
eigenvalues. We have to emphasize that for quarks in the adjoint
representation, the exact expressions for the anomalous dimensions \re{pair1}
coincide with those in the multi-color limit. At the same time, for quarks in
the fundamental, the baryon operator $B_{\rm f}(z_i)$ exists only for $N_c = 3$,
so that the large-$N_c$ counting is not applicable. The non-planar diagram
\ref{twoloopkernel}~(b) contributes to \re{pair2} and partially leads to
breaking of integrability in two loops.

Remarkably enough, the degeneracy of the eigenstates of the two-loop dilatation
operator $\mathbb{H}$ survives the conformal symmetry breaking. To understand
the phenomenon, one considers the dilatation operator $\mathbb{H}$ in the large
$\beta_0-$limit, $\lambda \beta_0={\rm fixed}$ and $\beta_0\to\infty$. In this
limit, the  $(n+1)-$loop Hamiltonian $\mathbb{H}$ is dominated by the
contribution of $\lambda(\lambda\beta_0)^n-$terms, which can be resummed. In
two loops, $\lambda^2\beta_0-$correction to \re{H-2-loop} comes from the second
term in \re{BetaInKernel} involving the kernel $\mathbb{H}_{jk}^{\scriptscriptstyle (1)}$.
Being combined with the one-loop kernel, the sum $\mathbb{H}_{jk}^{\scriptscriptstyle (0)}
+\lambda_{\scriptscriptstyle\rm R } ({\beta_{0}}/{c_{\scriptscriptstyle\rm R }}) \,
\mathbb{H}_{jk}^{\scriptscriptstyle (1)}$ is determined by the same integral
operator as $\mathbb{H}_{jk}^{\scriptscriptstyle (0)}$,
Eq.~\re{LightConeKernel}, provided that one substitutes
\be
\bar\alpha^{2 j_q - 1} \to \bar\alpha^{2 j_q - 1}\left[1 + \frac{\beta_0 g^2}
{16 \pi^2}\left( \frac53 + \ln\bar\alpha\right)\right].
\label{renorm}
\ee
Adding higher-loop $\lambda(\lambda\beta_0)^n-$corrections, one finds that the
$\ln\bar\alpha-$terms in the right-hand side of \re{renorm} exponentiate and
additively renormalize the bare quark conformal spin as $j_q \to j_q + \beta_0
g^2/(32 \pi^2)$ \cite{BelMul97}. As a result, the all-loop dilatation operator
$\mathbb{H}$, Eqs.~\re{DilatationOperator}, is given in the large$-\beta_0$ limit
by the one-loop expression \re{LOkernel} with the quark conformal spin $j_q=1$
replaced by its `renormalized' value
\be
\mathbb{H}^{(\beta_0\to\infty)} = \lambda\, \varphi(\ft{\beta_0 g^2} {16 \pi^2})
\cdot \mathbb{H}^{\scriptscriptstyle (0)}_{\phantom{i}}\big|_{j_q =1 +
\frac{\beta_0 g^2}{32 \pi^2}}\,,
\ee
with $\varphi(x) = \ft{(1+x)\Gamma(4+2x)}{6\Gamma(1-x)\Gamma^3(2+x)}$,
 $\lambda=\lambda_{\rm f}$ and $\lambda_{\rm a}$ for the operators
$\mathbb{B}_{\rm f}(z_i)$ and $\mathbb{B}_{\rm a}(z_i)$, respectively. The
operator $\mathbb{H}^{(\beta_0\to\infty)}$ inherits integrability of the one-loop
dilatation operator $\mathbb{H}^{\scriptscriptstyle (0)}$ and it can be mapped
into the Hamiltonian of the Heisenberg magnet of spin ${j_q =1+\beta_0 g^2/(32
\pi^2)}$. It possesses two conserved charges $q_2^{\scriptscriptstyle
(\beta_0\to\infty)}$ and $q_3^{\scriptscriptstyle (\beta_0\to\infty)}$ which are
given by the same expressions as before, Eq.~\re{charges}, where $L_{jk}^2$ is
given by \re{L2} with $j_q =1+\beta_0 g^2/(32 \pi^2)$. Thus, the $\beta_0-$terms
preserve integrability of the dilatation operator for both operators in
\re{B-operator} and do not lift the degeneracy of its spectrum.

For $\beta_0=0$, integrability holds to two loops only for adjoint quarks.
Indeed, introducing the operator $ \mathbb{Z} = ( \II + \mathbb{P}_{12} )
\mathbb{Z}^+/2 + ( \II - \mathbb{P}_{12} ) \mathbb{Z}^-/2 $ with $\mathbb{Z}^\pm$
defined in \re{PT}, it is straightforward to verify that the charges
\be
q_{n} = q_{n}^{\scriptscriptstyle (0)} + \lambda_{\rm a}
[\mathbb{Z},q_{n}^{\scriptscriptstyle (0)}] + \mathcal{O}(\lambda_{\rm
a}^2)\,,\quad (n=2,3)
\label{q-two-loop}
\ee
satisfy $[q_2,q_3]=[q_2,\mathbb{H}]=[q_3,\mathbb{H}]=0+\mathcal{O}(\lambda_{\rm
a}^2)$ provided that the spectrum of $\mathbb{H}$ is degenerate,
$\gamma^{\scriptscriptstyle (1)+}_q  = \gamma^{\scriptscriptstyle (1)-}_q$.
Notice that the two-loop dilatation operator $\mathbb{H}$ in the $\overline{\rm
MS}-$scheme does not respect the conformal symmetry even for $\beta(g)=0$,
$[\mathbb{H},q_2^{\scriptscriptstyle (0)}]\neq 0$. The reason for this is that
the dimensional regularization inevitably breaks conformal symmetry since
$\beta(g) = (d - 4) g/2$ is non-zero in $d-$dimensions and it propagates into the
anomaly as $d\to 4$. The anomaly can be cured by a finite renormalization of the
$\mathbb{B}-$operator, $\mathbb{B}_U(z_i) =
[\mathbb{U}\!\cdot\!\mathbb{B}](z_i)$, leading to the dilatation operator,
$\mathbb{H}_U=\mathbb{U}\!\cdot\!\mathbb{H}\!\cdot\!\mathbb{U}^{-1}$~\cite{BelMul98}.
This does not affect the spectrum of the anomalous dimensions but modifies the
conserved charges $q_2, q_3$ and the eigenfunctions $\Psi_q(z_i)$. The operator
$\mathbb{U}$ defines a scheme in which the conformal symmetry of the dilatation
operator is restored for $\beta(g)=0$. Such scheme is not unique and can be
defined for our best convenience. Choosing $\mathbb{U}=1-\lambda\, \mathbb{Z}+
\mathcal{O}(\lambda^2)$ with the same $\mathbb{Z}$ as in \re{q-two-loop}, one
can define a scheme in which the charges do not receive radiative corrections
$(q_n)_U =\mathbb{U}q_n\mathbb{U}^{-1}= q_n^{\scriptscriptstyle (0)}+\mathcal{O}
(\lambda^2)$ while the two-loop correction to the dilatation operator $\mathbb{H}_U
=\mathbb{H}^{\scriptscriptstyle (0)} + \lambda\, \mathbb{H}_U^{\scriptscriptstyle (1)}
+\mathcal{O}(\lambda^2)$ with $\mathbb{H}_U^{\scriptscriptstyle (1)}=
\mathbb{H}^{\scriptscriptstyle (1)}_{\phantom{i}} - [\mathbb{Z},
\mathbb{H}^{\scriptscriptstyle (0)}_{\phantom{i}} ]$ commutes with the one-loop
kernel, $[\mathbb{H}^{\scriptscriptstyle (0)}, \mathbb{H}_U^{\scriptscriptstyle (1)}]=0$.

So far we discussed the three-quark operators \re{B-operator}. As was already
mentioned, at $N_f=1$ the operator $\mathbb{B}_a(z_i)$ is a component of the
supermultiplet in the $\mathcal{N}=1$ SYM. Supersymmetry allows one to extend the
two-loop integrability to remaining components of the supermultiplet including
three-gluon operator of the maximal helicity \cite{BraDerMan98}. Going over to
$\mathcal{N} = 2$ and $\mathcal{N} = 4$ SYM theories, one finds that the only
modification one has to make is to add to the dilatation operator \re{H-2-loop}
the contribution of two-loop diagrams with, respectively, two and six
real scalars inside the loops of two-particle kernels but not in the external
lines since the maximal-helicity operators carrying the maximal $R-$charge mix
only among themselves.  The scalars modify the constant $\Gamma^{\scriptscriptstyle
(1)}$ in \re{H-2-loop} and induce an additional contribution to the two-particle
kernels $\mathbb{H}_{jk}$ proportional to $\mathbb{H}_{jk}^{\scriptscriptstyle
(0)}$ and $\mathbb{H}_{jk}^{\scriptscriptstyle (1)}$. Since these terms preserve
integrability, the two-loop integrability found above is promoted to the
$\mathcal{N} = 2$ and $\mathcal{N} = 4$ SYM theories.

\end{document}